\documentclass[twocolumn, aps, prb,amsmath,amssymb,citeautoscript,longbibliography]{revtex4-1}

\pdfoutput=1
\usepackage{natbib}
\usepackage[english]{babel}
\usepackage{letltxmacro}
\usepackage{latexsym}
\LetLtxMacro{\ORIGselectlanguage}{\selectlanguage}
\makeatletter
\DeclareRobustCommand{\selectlanguage}[1]{%
  \@ifundefined{alias@\string#1}
    {\ORIGselectlanguage{#1}}
    {\begingroup\edef\x{\endgroup
       \noexpand\ORIGselectlanguage{\@nameuse{alias@#1}}}\x}%
}
\newcommand{\definelanguagealias}[2]{%
  \@namedef{alias@#1}{#2}%
}
\makeatother
\definelanguagealias{en}{english}
\definelanguagealias{English}{english}
\usepackage{graphicx}
\usepackage{amsmath}
\usepackage{amsfonts}
\usepackage{amssymb,bbding}
\usepackage{bm}
\usepackage[title]{appendix}
\usepackage{color}
\usepackage[percent]{overpic}
\usepackage{soul} 
\usepackage{amssymb}
\usepackage{wasysym}
\usepackage{dsfont}
\usepackage{float}
\usepackage{braket}
\usepackage{cancel}
\usepackage{comment}

\usepackage{enumitem}
\usepackage[dvipsnames]{xcolor}
\usepackage{graphicx}

\usepackage{mathtools}
\usepackage{comment}
\usepackage{relsize}
\usepackage{dsfont} 
\usepackage{commath}
\usepackage{titlesec}

\usepackage{hyperref}
\hypersetup{
    unicode=false,          
    pdftoolbar=false,       
    pdfmenubar=true,        
    pdffitwindow=false,     
    pdfstartview={FitH},    
    pdftitle={},            
    pdfauthor={Authors},    
    pdfsubject={},          
    pdfcreator={},          
    pdfproducer={},         
    pdfkeywords={QAOA}{interaction distance},
    pdfnewwindow=true,      
    colorlinks=true,        
    linkcolor=black,        
    citecolor=blue,         
    filecolor=magenta,      
    urlcolor=blue           
}

\usepackage{cleveref}

\usepackage[utf8]{inputenc}
\usepackage[export]{adjustbox}

\setlist[itemize]{leftmargin=*}
\setlist[enumerate]{leftmargin=*}

\crefformat{equation}{Eq.~(#2#1#3)}
\Crefformat{equation}{Eq.~(#2#1#3)}
\crefrangeformat{equation}{Eqs.~(#3#1#4--#5#2#6)}
\crefname{figure}{Fig.}{Fig.}

\DeclareMathOperator*{\trace}{Tr}

\newcommand*{\gmfld}{\mathcal{F}}
\newcommand*{\df}{D_\gmfld}

\begin{document}

\title{Quantifying the efficiency of state preparation via quantum variational eigensolvers}

\author{Gabriel Matos$^1$, Sonika Johri$^2$, and Zlatko Papi\'c$^1$}
\affiliation{$^1$School of Physics and Astronomy, University of Leeds, Leeds LS2 9JT, United Kingdom}
\affiliation{$^2$IonQ Inc., College Park, MD 20742, USA}

\date{\today}

\begin{abstract}

Recently, there has been much interest in the efficient preparation of complex quantum states using low-depth quantum circuits, such as Quantum Approximate Optimization Algorithm (QAOA). While it has been numerically shown that such algorithms prepare certain correlated states of quantum spins with surprising accuracy, a systematic way of quantifying the efficiency of QAOA in general classes of models has been lacking. Here, we propose that the success of QAOA in preparing ordered states is related to the interaction distance of the target state, which measures how close that state is to the manifold of all Gaussian states in an arbitrary basis of single-particle modes. 
We numerically verify this for several examples of non-integrable quantum models, including Ising models with two- and three-spin interactions and the cluster model in an external field.
Our results suggest that the structure of the entanglement spectrum, as witnessed by the interaction distance, correlates with the success of QAOA state preparation, and that this correlation also contains information about different phases present in the model. We conclude that QAOA typically finds a solution that perturbs around the closest free-fermion state.

\end{abstract}

\maketitle{}

\section{Introduction}
\label{sec:introduction}

In recent years,  algorithms involving a hybrid quantum-classical procedure for cost function minimization have attracted much attention~\cite{peruzzo_variational_2014, farhi_quantum_2014}. Among these is the Quantum Approximate Optimization Algorithm (QAOA), which employs an alternating operator ansatz for solving  optimization problems that are mappable to the problem of finding the ground state of a classical Ising-type Hamiltonian~\cite{farhi_quantum_2014}.  These ansatz circuits are of great interest since they have been shown to successfully approximate or even exactly prepare states at remarkably \emph{low} circuit depths. This makes them amenable to implementation using the currently available ``noisy intermediate scale" quantum computers~\cite{preskill_quantum_2018}, potentially enabling useful applications in the near future before full-fledged quantum computers, featuring robust error correction, may become operational. It has also been proven that QAOA circuits can perform universal quantum computation for certain classes of Hamiltonians~\cite{lloyd_quantum_2018,morales_universality_2019}.

QAOA was originally proposed to tackle classical optimization problems, such as the MaxCut problem~\cite{farhi_quantum_2019}, and several others~\cite{vikstal_applying_2019, hadfield_quantum_2019, oh_solving_2019, sundar_quantum_2019}.  More recently, it has been pointed out that QAOA could also serve as a tool for exactly preparing \emph{quantum} many-body states, such as the GHZ state, the ground state of the Ising model at the critical point, for both short-range~\cite{ho_efficient_2019} and long-range interactions~\cite{ho_ultrafast_2019,wauters_polynomial_2020},  the ground state of the toric code~\cite{ho_efficient_2019}, the ground state of the two-dimensional Hubbard model~\cite{vogt_preparing_2020}, and the thermofield double states~\cite{zhu_variational_2019, wu_variational_2019, premaratne_engineering_2020}. In this paper we focus on the latter type of applications of QAOA in the context of preparing ground states of non-integrable quantum Hamiltonians.

Some analytical results on state preparation in the \emph{classical} Ising model have been established. It was shown that in one dimension, the uniform nearest-neighbor Ising model in the absence of a magnetic field can be reduced to a system of pseudospins~\cite{wang_quantum_2018}. This reduction was used to prove a conjecture~\cite{farhi_quantum_2014}  that the ground state of this model can be prepared using a circuit with depth linear in the system size~\cite{wang_quantum_2018}, and the associated bounds on the best attainable fidelity for QAOA circuits below this depth were derived~\cite{mbeng_optimal_2019}. At the same time, a deeper understanding of QAOA that would, for example, allow for the systematic construction of a circuit to prepare a given quantum state, and to predict the circuit depth needed to reach a certain accuracy, is still lacking.

In its original formulation, QAOA involved the alternating application of a ``mixer" Hamiltonian and a ``problem" Hamiltonian~\cite{farhi_quantum_2014,hadfield_quantum_2019}, where different choices for the mixer Hamiltonian can be made \cite{farhi_quantum_2014, wang_xy-mixers_2020, bartschi_grover_2020}. A different approach is to split the problem Hamiltonian into a number of terms and alternate between the application of those; in this way, QAOA can be seen as the digitization plus Trotter splitting of a quantum annealing protocol~\cite{mbeng_optimal_2019}. This justifies that, as the circuit depth increases, states can be prepared with increasing accuracy. It does not explain, however, why some states can be prepared with very high accuracy using low circuit depths.  Thus, there remain open questions about the inner workings of QAOA. Some of the difficulties in developing a deeper understanding of QAOA, as well as its numerical implementations, stem from the fact that the optimization landscape is generally  riddled with local minima and other issues~\cite{cerezo_cost-function-dependent_2020, bukov_broken_2018, day_glassy_2019}. Different techniques, such as alternative optimization methods~\cite{yamamoto_natural_2019, sweke_stochastic_2019, shaydulin_multistart_2019}, modifications to the cost function\cite{premaratne_engineering_2020, li_quantum_2020}, or machine learning techniques~\cite{garcia-saez_quantum_2019, khairy_learning_2019, khairy_reinforcement-learning-based_2019, alam_accelerating_2020, yao_policy_2020},  have attempted to address these problems. Heuristics for producing a starting set of optimization angles have also been developed~\cite{pichler_quantum_2018, zhou_quantum_2019, pagano_quantum_2020}.

In this paper, we address the problem of using QAOA to prepare the ground state of some quantum Hamiltonian $H$ that depends on one or more tunable parameters $H(h_1,h_2,\ldots)$, such that $H$ is non-integrable for general values of $h_1$, $h_2$, etc. We are interested in predicting the relative success of QAOA state preparation across the phase diagram defined by the parameters $h_1$, $h_2, \ldots$; moreover, our aim is to relate the success of  preparation to some physical property of the target state.  It has been argued that the von Neumann (entanglement) entropy of the target state may determine the quality of the variational approximation at very low circuit depths~\cite{BravoPrieto2020scalingof}. However, at these depths, the states prepared by QAOA are generally still far from the target state.  We find that, ultimately, the quality of QAOA state preparation correlates with the property called \emph{interaction distance}\cite{turner_optimal_2017}. The latter can be evaluated from the eigenvalue spectrum of the system's (reduced) density matrix. Our findings are numerically supported by examples of non-integrable quantum Ising models.

The remainder of this paper is organized as follows. Secs.~\ref{sec:QAOA} and  \ref{sec:df} contain a brief overview of the QAOA variational ansatz and interaction distance, respectively. In Sec.~\ref{sec:ising_model}, we introduce an alternating operator protocol for the Ising model in transverse and longitudinal fields,   and we demonstrate that the success of the QAOA ground-state preparation correlates with its interaction distance. We show that the \emph{slope} of this cross-correlation can be used to identify the existence of different phases  in the model. In Sec.~\ref{sec:relation} we provide analytical arguments for the numerically-observed correlation between QAOA and interaction distance, while in Sec.~\ref{sec:landscape} we analyze the optimization landscape for these models. Our conclusions are presented in Sec.~\ref{sec:conclusion}, while Appendices contain generalizations of our results. In particular, Appendix~\ref{sec:three-spin} contains the results for a variant of the  Ising model which features interactions between nearest-neighbour triplets of spins. This model has a critical line in the universality class of the Potts model~\cite{francesco_conformal_2012}, and its ground state is much harder to prepare than that of the quantum Ising model. On the other hand, Appendix~\ref{sec:cluster} contains results for the model which realises the so-called cluster state~\cite{Briegel2001}, which is of importance in measurement-based quantum computation~\cite{Nielsen2006} and also in symmetry-protected topological phases of matter~\cite{Pollmann2010, Fidkowski2011, Chen2011,Schuch2011}. This model displays a critical line in its phase diagram when placed in an external magnetic field, similar to the AFM model, and we demonstrate similar correlation between the interaction distance and QAOA preparation of its ground state.

\section{Quantum Approximate Optimization Algorithm}
\label{sec:QAOA}

Various names for quantum-classical variational algorithms have been proposed in the literature, depending on the context and the specific implementation~\cite{peruzzo_variational_2014, mcclean_theory_2016,farhi_quantum_2014, reiner_finding_2019, ho_efficient_2019,vogt_preparing_2020}.  Among the first such algorithms is  the Variational Quantum Eigensolver (VQE) -- proposed in the context of quantum chemistry~\cite{peruzzo_variational_2014, mcclean_theory_2016} -- for preparing  approximate eigenstates and calculating eigenvalues of a given Hamiltonian. The QAOA~\cite{farhi_quantum_2014} introduced the alternating operator ansatz, which we review below.  We refer to the general class of variational quantum-classical algorithms based on the alternating operator ansatz simply as QAOA, with the understanding that it can be seen as a specialization of VQE for this particular class of ans\"{a}tze.

As previously mentioned, QAOA is a variational algorithm based on a  classical optimization routine which performs a minimization over a parametrized family of quantum circuits. The goal of this minimization is to find the circuit which, starting from some initial state $|\psi_\mathrm{initial}\rangle$, best prepares the target quantum state, $|\psi_\mathrm{target}\rangle$. This family of quantum circuits is defined by a set of operators $H_1, H_2, \ldots, H_M$, and takes the alternating operator ``bang-bang" form~\cite{yang_optimizing_2017}, defined by the unitary
\begin{eqnarray}\label{qaoa_protocol}
\nonumber	U(\bm{\theta}) &=& e^{-i \theta_{p,1} H_1} e^{-i \theta_{p,2} H_2} \ldots e^{-i \theta_{p, M} H_M} \\
	&& \dots e^{-i \theta_{1, 1} H_1} e^{-i \theta_{1, 2} H_2} \ldots e^{-i \theta_{1, M} H_M}.
\end{eqnarray}
The circuit is parameterized by the set of variational angles $\bm \theta = (\theta_{1,1},...,\theta_{p,M})$. Note that in most of the paper (with the exception of Sec.~\ref{sec:landscape}) we do not place restrictions on the total ``time" taken by the protocol, $\sum_{i=1}^p \sum_{j=1}^{M}\theta_{i,j}$. If this total time  is fixed, it has been argued that the optimal protocol is a hybrid consisting of ``bang-bang" near the beginning and end of the evolution, with smooth annealing in between~\cite{brady_optimal_2020}. 

\begin{figure}[ht]
	\centering
	\includegraphics[width=\linewidth]{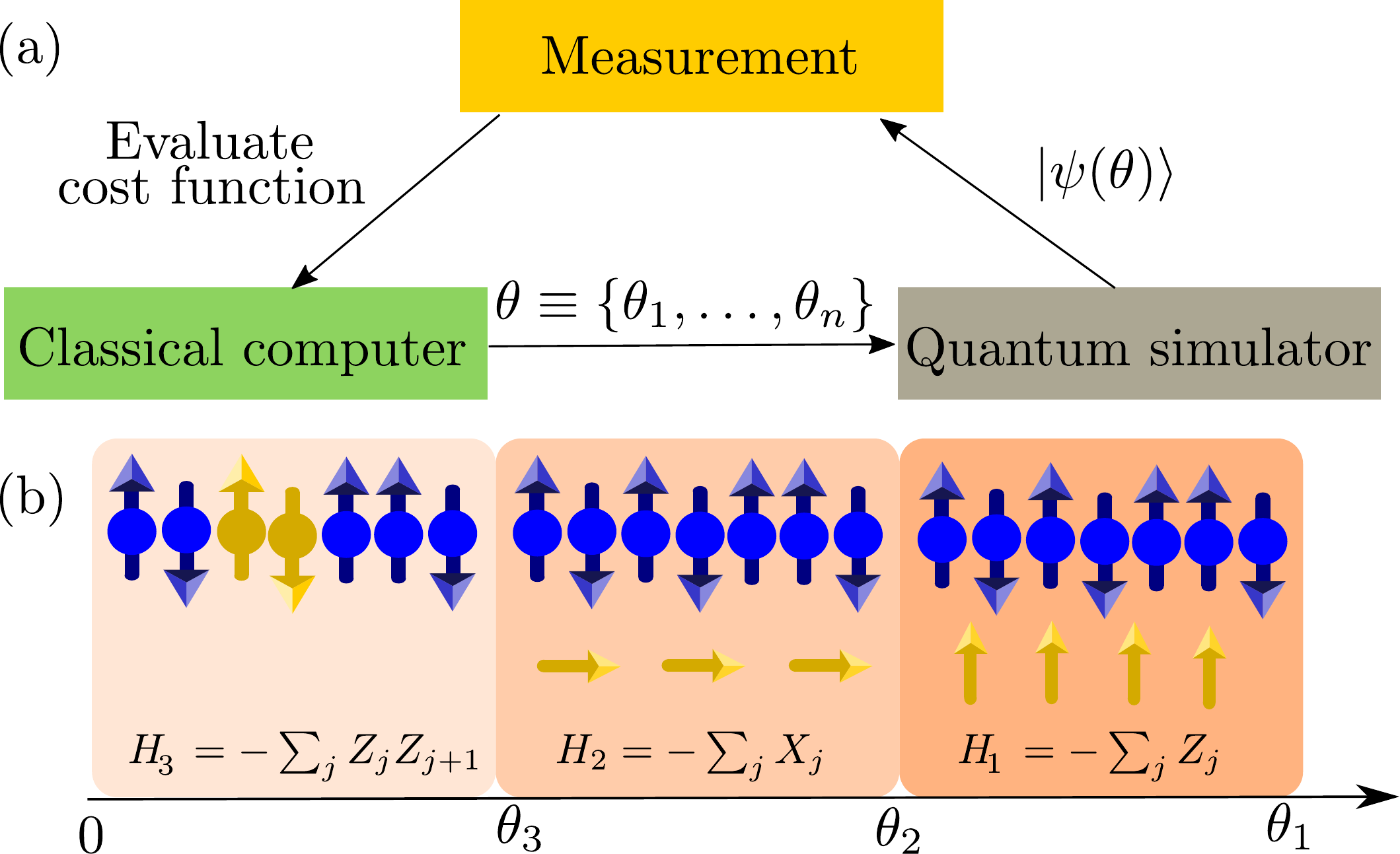}
	\caption{(a) A schematic illustrating a variational quantum-classical optimization routine. The optimization involves $n$ parameters $\theta_i$, where $i=1,2,\ldots,n$. For the ansatz in Eq.~(\ref{qaoa_protocol}), $n=pM$.
	(b) The $M=3$-step QAOA algorithm for the preparation of the ground state of the Ising model in both transverse and longitudinal fields, discussed in Sec.~\ref{sec:ising_model}.}
	\label{fig:qaoa}
\end{figure}

A sketch of the QAOA protocol is given in Fig.~\ref{fig:qaoa}(a). The  algorithm starts with some chosen initial state  $\ket{\psi_\mathrm{initial}}$ and an initial set of values for the circuit parameters.
The initial state $|\psi_\mathrm{initial}\rangle$ is, in principle, arbitrary, but it should be sufficiently easy to prepare (e.g., a product state or some low-entangled state). The target state $\psi_\mathrm{target}$ is often assumed to be the ground state of some Hamiltonian $H$, sometimes called the ``problem Hamiltonian''.  As mentioned previously, there is freedom in the choice of the operators $\{H_j\}_{j \in \{1,...,M\}}$. Unlike the original formulation~\cite{farhi_quantum_2014}, for the problems considered in this paper, we choose the operators by splitting the problem Hamiltonian as $H = \sum_{i=1}^M h_i H_i$ for $M=3$ (see Sec.~\ref{sec:ising_model} below), where the $h_i$ are the tunable parameters for the models we study. 

After preparing the initial state, the simulator performs the quantum evolution
\begin{eqnarray}\label{vqcs_protocol}
\nonumber	\ket{\psi(\bm{\theta})} &=& U(\bm{\theta}) \ket{\psi_\mathrm{initial}}.\quad
\end{eqnarray}
After the state $\ket{\psi(\bm{\theta})}$ is obtained, a cost function is measured. In what follows, we assume all states to be normalized. The cost function may be defined as the expectation value of the energy
\begin{eqnarray}
	E \equiv \braket{\psi(\bm{\theta})|H|\psi(\bm{\theta})}.
\end{eqnarray}
It is often more convenient to use the rescaled relative energy~\cite{mbeng_quantum_2019, pagano_quantum_2020} 
\begin{eqnarray}\label{eq:relative_energy}
	\epsilon \equiv \frac{\braket{\psi(\bm{\theta})|H|\psi(\bm{\theta})}- E_{\min}}{E_{\max} - E_{\min}},
\end{eqnarray}
where $E_\mathrm{min}$, $E_\mathrm{max}$ are the extremal eigenenergies in the spectrum of $H$. The relative energy $\epsilon$ is  bounded between $0$ and $1$, such that $\epsilon=0$ corresponds to finding an exact ground state. 
Alternatively, if $|\psi_\mathrm{target}\rangle$ is known, the cost function can be taken to be the quantum infidelity,
\begin{eqnarray}
	1 - f \equiv 1- |\langle \psi_\mathrm{target} | \psi(\bm{\theta}) \rangle|^2,
\end{eqnarray}
which is similarly bounded between 0 and 1. Although evaluating fidelity in experiment is impractical or even impossible, it is often useful in numerical simulations.

Once the value of the cost function is measured, it is passed back to the optimization algorithm running on the classical computer. This algorithm returns a new set of angles, which are passed again to the quantum simulator, and the process repeats itself until the optimization algorithm running on the classical computer halts.

In Ref.~\onlinecite{ho_efficient_2019} it was observed that the ground state of the transverse-field Ising model with periodic boundary conditions could be prepared \emph{exactly} (i.e., with $f=1$) in precisely $p=N/2$ steps, where $N$ is the total number of spins. This was done using the same $M=2$-operator QAOA protocol originally proposed for the MaxCut problem~\cite{farhi_quantum_2014}. This was a surprising result, given that the ground state of the Ising model can be very complex depending on the magnitude of the transverse magnetic field. For instance, at the critical value of the field, the excitation gap goes to zero and the ground state displays logarithmically diverging von Neumann entropy (VNE) of entanglement as a function of subsystem size~\cite{calabrese_entanglement_2004}.  This example demonstrates that the success of QAOA protocol at that circuit depth is not determined by the VNE of the target state. One of the main results of the present paper is to show  that a different quantum-information measure called \emph{interaction distance}~\cite{turner_optimal_2017} serves as an error-metric for the quality of QAOA state preparation. In the following section, we briefly introduce and review the properties of interaction distance (see also Ref.~\onlinecite{pachos_quantifying_2018}).

\section{Interaction distance}
\label{sec:df}

Given some density matrix $\rho$, the \emph{interaction distance}~\cite{turner_optimal_2017} of $\rho$ is defined as
\begin{align} \label{eq_df}
	\df(\rho) := \min_{\sigma \in \mathcal{F}} \frac{1}{2} \trace\left( \sqrt{(\rho - \sigma)^2} \right),
\end{align}
where $\mathcal{F}$ is the manifold of Gaussian density matrices $\sigma$,
\begin{align}\label{eq:gaussian_manifold}
	\mathcal{F} &:= \{\sigma = \frac{1}{Z} e^{-H},  Z = \trace e^{-H}, \text{$H$ is quadratic} \}.
\end{align}
Here, $H$ being quadratic means that it is a free-particle Hamiltonian, e.g., in second quantization,  $H = c^\dagger h c$ for some matrix $h$ and some set of creation and annihilation operators $\{c_j^\dagger \}$, $\{ c_j\}$, with $j \in \{1, ..., N\}$.

$D_\mathcal{F}$, as defined in Eq.~(\ref{eq_df}), measures distinguishability between a given density matrix $\rho$ and the set of all free-particle density matrices, $\sigma$. Physically, the density matrix $\rho$ can represent a thermal state of the system, in which case it is the standard Boltzmann-Gibbs density matrix when the system is in thermodynamic equilibrium at some temperature $\beta=1/T$. On the other hand,  $\rho$ can also be a reduced density matrix which describes the subsystem $A$ of a larger system in the pure state $|\psi\rangle$. In this case, $\rho$ is  obtained as the partial trace $\rho_A := \mathrm{Tr}_{\bar A} \; |\psi\rangle \langle \psi|$ over the degrees of freedom of the subsystem $\bar A$ complementary to the subsystem $A$. 

The reduced density matrix is  useful for characterizing properties of $|\psi\rangle$, such as the entanglement entropy of $A$,
\begin{eqnarray}
S_{\mathrm{VNE}} = - \mathrm{Tr} \; \rho_A \ln \rho_A.
\end{eqnarray}
Since $\rho_A$ is readily available in numerical simulations, in what follows we focus on $D_\mathcal{F}$ evaluated with respect to the reduced density matrix of the model's ground state.

There is a crucial simplification in evaluating $D_\mathcal{F}$ as written in Eq.~(\ref{eq_df}), which was shown  in Ref.~\onlinecite{turner_optimal_2017} using results from ~Ref.~\onlinecite{markham_quantum_2008}. The minimization over $\mathcal{F}$ is equivalent to
\begin{eqnarray}\label{eq:df_sim}
	\df(\rho) = \min_{ \{\epsilon_j \}} \frac{1}{2} \sum_{k} \abs{\rho_k - \sigma_k(\bm{\epsilon})},
\end{eqnarray}
where $\rho_k$ denote the eigenvalues of  $\rho$ in descending order (normalized such that $\sum_k \rho_k = 1$), and 
\begin{align} \label{eq_df_reduced}
	\sigma_k(\bm{\epsilon}) = \frac{1}{Z} e ^{-\sum_{j} \epsilon_j n_j^{(k)}},
\end{align}
where $n_j^{(k)} \in \{0, 1\}$ is the occupancy number on the $j$th site of the $k$th element of a Fock basis with the energy $\epsilon_j$. The normalization $Z$ ensures that $\sum_k \sigma_k = 1$, and we assume that $\sigma_k$ are in  the same (descending) order as $\rho_k$, which is necessary to achieve a minimum in Eq.~(\ref{eq:df_sim})~\cite{markham_quantum_2008}. 

The utility of Eq.~(\ref{eq:df_sim}) is that the value of $D_\mathcal{F}(\rho)$ can thus be determined solely from the information of the spectrum $\{-\log\rho_k\}$, also known as the ``entanglement spectrum"~\cite{li_entanglement_2008}. Comparing Eq.~(\ref{eq_df}) with Eq.~(\ref{eq:df_sim}), we see that the minimization over all matrices $\sigma\in\mathcal{F}$ was traded for a minimization over scalars $\{\epsilon_j\}$. The latter is a much simpler optimization problem, as the number of $\epsilon$ parameters is expected to scale linearly with the system size. Thus, the problem becomes numerically tractable, as the computational complexity is only polynomial in system size $N$ once the spectrum $\{\rho_k\}$ is known~\cite{turner_optimal_2017}. 

Note that $D_\mathcal{F}$ is strictly bounded $0\leq D_\mathcal{F} \leq 1$~\cite{meichanetzidis_free-fermion_2018}, and states that have $D_\mathcal{F}=0$ can be expressed as Gaussian states in terms of some free-particle modes as in Eq.~(\ref{eq_df_reduced}). This is, of course, true for unentangled (product states) in the computational basis, but it is also the case for certain entangled states. An example is the ground state of the Ising model in the transverse field, as discussed in the following section. Interestingly, unlike the lower bound, $D_\mathcal{F}$ does not seem to saturate its upper bound -- it was conjectured that $D_\mathcal{F} \leq 3-2\sqrt{2}$~\cite{meichanetzidis_free-fermion_2018}. Physical states that realize this upper bound of $D_\mathcal{F}$ were identified as ground states of certain types of parafermion chains~\cite{meichanetzidis_free-fermion_2018}. These states do not have a particularly high value of VNE, but the structure of their entanglement spectrum is as distinct as possible from that of free fermions, in the sense of Eq.~(\ref{eq:df_sim}).

\section{Preparing the ground state of the non-integrable quantum Ising model}
\label{sec:ising_model}

In this section we present our main findings on the correlation between the success of QAOA state preparation and the interaction distance of the target state. As a toy model, we consider the one-dimensional quantum Ising model in the presence of both transverse and longitudinal fields,
\begin{align} \label{eq:ising_hamiltonian}
	H &= - \sum_{i = 1}^{N} (\pm 1) Z_i Z_{i+1} - h_x \sum_{i = 1}^{N} X_i - h_z \sum_{i = 1}^{N} Z_i ,
\end{align}
where $X_i$ and $Z_i$ are the standard Pauli matrices on site $i$, and we assume periodic boundary conditions (PBCs) by identifying sites $j+N \equiv j$. The model is either ferromagnetic (FM) or antiferromagnetic (AFM) depending on whether the coupling of the first term is chosen to be $+1$ or $-1$, respectively. The Ising models in Eqs.~(\ref{eq:ising_hamiltonian}) serve as a useful laboratory for studying a number of phenomena in condensed matter physics~\cite{dutta_quantum_2015, coldea_quantum_2010, morris_hierarchy_2014}.

\begin{figure}[ht]
	\centering
	\includegraphics{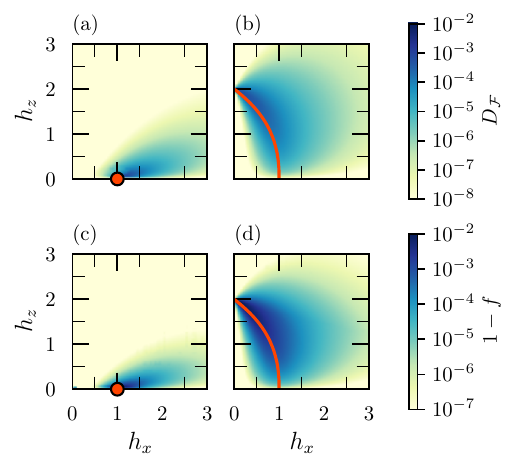}
	\caption{Top row: Interaction distance for the ground state of FM (a) and AFM (b) model in Eq.~(\ref{eq:ising_hamiltonian}) as a function of fields $h_x$ and $h_z$. Bottom row: the infidelity $1-f$ of the QAOA protocol for the FM (c) and AFM model (d). See text for the details of the QAOA protocol. All data is for system size $N=8$ using PBCs. Red dot denotes the critical point of the FM model, while the red line is the critical line in the AFM model according to Ref.~\onlinecite{ovchinnikov_antiferromagnetic_2003}.
	}
	\label{fig:4hmap}
\end{figure}
The properties of the ground state of the model in Eq.~(\ref{eq:ising_hamiltonian}) are insensitive to the sign of the FM/AFM coupling in the absence of  longitudinal field $h_z$. However, once $h_z>0$, the phase diagram is substantially different for the two models. The FM model has a critical point at $(h_x, h_z) = (1,0)$, while  the AFM  model has a critical line connecting the point $(h_x, h_z) = (1,0)$ with the point $(h_x, h_z) = (0,2)$. The critical line is not known analytically, but it has been determined numerically using density-matrix renormalization group simulations in Ref.~ \onlinecite{ovchinnikov_antiferromagnetic_2003}. In both cases, the limit of purely transverse field ($h_z = 0$) is particularly important. Along this line, the Hamiltonian is diagonal when written in terms of free fermions  after performing a combination of the Jordan-Wigner and Bogoliubov transforms~\cite{pfeuty_one-dimensional_1970}.

The phase diagrams of the FM and AFM models diagnosed by the value of  $D_\mathcal{F}$  in their ground state are shown in Fig.~\ref{fig:4hmap}(a)-(b), respectively. The ground state of the Hamiltonian in Eq.~(\ref{eq:ising_hamiltonian}) is obtained numerically using exact diagonalization, and its entanglement spectrum is computed by partitioning the system into two equal halves. From the entanglement spectrum, $D_\mathcal{F}$ is evaluated by numerical optimization following Eq.~(\ref{eq:df_sim})~\footnote{\href{https://bitbucket.org/cjturner/esfactor/}{https://bitbucket.org/cjturner/esfactor/}}. For both models, $D_\mathcal{F}$ is found to be zero (to machine precision) when $h_z=0$, regardless of the value $h_x$. Away from this line, $D_\mathcal{F}$ is a sensitive indicator of interaction effects and changes by many orders of magnitude depending on the location in the phase diagram. For example, in the FM model, $D_\mathcal{F}$ exhibits a sharp peak just off the free Ising critical point, $(h_x=1,h_z=0)$. While the Ising critical point is described by the free Ising conformal field theory~\cite{francesco_conformal_2012} and thus it has $D_\mathcal{F}=0$, the properties of this CFT change dramatically once $h_z$ field is introduced~\cite{zamolodchikov_integrals_1989}. This is consistent with the fermionic picture, where the $h_z$ field introduces  long-range interaction between fermions after the Jordan-Wigner transformation, which makes the system's ground state highly interacting. Somewhat surprisingly, away from the critical point, the value of $D_\mathcal{F}$ sharply decays to values as low as $\sim 10^{-7}$, even though the interaction is comparable in magnitude to other terms in the Hamiltonian. This implies that there are large regions of the phase diagram where the ground state of the system is effectively free-fermion-like, even though the Hamiltonian itself is ``interacting".
On the other hand, the AFM model features a critical line that extends from the free Ising critical point $(h_x=1,h_z=0)$. While $D_\mathcal{F}=0$ at $(h_x=1,h_z=0)$, the value of interaction distance progressively increases along the critical line towards the interior of the phase diagram -- see Fig.~\ref{fig:4hmap}(b). 

Next, we explore how to prepare the ground state of Eq.~(\ref{eq:ising_hamiltonian}) using QAOA for arbitrary values of fields $h_x$ and $h_z$. To this end, we have found it necessary to employ a $M=3$-step QAOA protocol from Eq.~(\ref{vqcs_protocol}) with the operators
\begin{eqnarray}
	H_1 &=& - \sum_{i = 1}^{N} Z_i, \label{eq:h3} \\
	H_2 &=& - \sum_{i=1}^{N} X_i, \label{eq:h2} \\
	H_3 &=& - \sum_{i = 1}^{N} Z_i Z_{i+1}, \label{eq:h1}
\end{eqnarray}
which satisfy $\pm H_3 + h_x H_2 + h_z H_1=H$, see the illustration in Fig.~\ref{fig:qaoa}(b). The initial state of the protocol is taken to be the ground state of $H_2$, i.e., all spins polarized along $x$-direction, $|\psi_\mathrm{init}\rangle = |\rightarrow\ldots \rightarrow\rangle$.
Since Pauli matrices are involutions, these operators $H_j$ satisfy $e^{-i(\theta + \pi) H_j} = \pm e^{-i\theta H_j}$, so in what follows we restrict the representation of $\bm{\theta}$ angles  to $[0,\pi[$ interval. We further restrict the angle $\theta_{p,3}$ associated with $H_3$ to the $[0, \frac{\pi}{2}[$ interval, since 
\begin{eqnarray}
\nonumber \prod_i^N e^{i(\theta +\frac{\pi}{2}) Z_i Z_{i+1}} &\propto & \prod_{i=1}^N Z_i Z_{i+1} \prod_{i=1}^N e^{i \theta Z_i Z_{i+1}} \\
&\propto&  \prod_{i=1}^N e^{i \theta Z_i Z_{i+1}}.
\end{eqnarray}

The initial guesses for the angles were determined sequentially as $p$ is increased, following the method in Appendix B1 of Ref.~\onlinecite{zhou_quantum_2019}. For minimizations involved in both QAOA and $\df$ we use a basinhopping algorithm with a Metropolis acceptance criterion~\cite{wales_global_1997}, as implemented in the Python package \texttt{scipy.optimize.basinhopping}. This is a global strategy that performs multiple minimizations, taking as the initial condition for the next minimization the stochastically perturbed result of the previous one. This allows us to avoid the local minima associated with the rugged landscapes of both QAOA and $\df$, as discussed further in Sec.~\ref{sec:landscape}. This, however, was not enough to completely eliminate local minima, and all the data presented here required two additional rounds of minimization. Each of these consisted in running the basinhopping algorithm across the phase diagram again, this time using as initial value for each point the optimal values of each of the adjacent points from the previously obtained data, and keeping the minimum value found.

Note that our protocol in Eqs.~(\ref{eq:h3})-(\ref{eq:h1}) is a generalization of the one considered in Ref.~\onlinecite{ho_efficient_2019}, which was restricted to the purely transverse field ($h_z=0$) and made use of a $M=2$-step ansatz with only $H_2$ and $H_3$. In that case, both the Hamiltonian and the protocol conserve the total fermion parity, generated by $P=\prod_i X_i$. This symmetry is broken once the $z$-field is introduced and the ground state acquires a non-zero magnetization $\braket{\psi|\sum_i Z_i |\psi} \neq 0$. While it is easy to come up with a two-step protocol that does not conserve parity, we have not been able to find one that accurately prepares the ground state for general values of $(h_x,h_z)$, thus we introduced a third operator into the QAOA protocol.

In Figs.~\ref{fig:4hmap}(c)-(d) we present results of the QAOA protocol across the phase diagram $(h_x,h_z)$. The color scale in Fig.~\ref{fig:4hmap}(c)-(d) shows the infidelity $1-f$ obtained after fixed $p=\frac{N}{2}$ steps of QAOA. We observe that this metric of ground state preparation looks remarkably similar to the behavior of $D_\mathcal{F}$ in Figs.~\ref{fig:4hmap}(a)-(b). In particular, we recover $f=1$ when $h_z=0$~\cite{ho_efficient_2019}, while the QAOA no longer finds an exact ground state when $h_z>0$. Nevertheless, it approximates the ground state very closely when $D_\mathcal{F}$ is small. Once again, it is easy to see that in this case there is no clear relation between QAOA's $1-f$ and the VNE of the ground state. For example, in the FM model, the VNE should be largest at the critical point; further, as adding $h_z$ opens a gap in the spectrum, increasing this parameter should reduce the VNE, as its scaling changes from logarithmic divergence with system size to an area law. However, from the point of view of QAOA, we find precisely the opposite: it is harder to prepare the state with some small amount of $h_z$ compared to $h_z=0$. 

Examining the optimal angles found at each point of the phase diagram of both the FM/AFM Ising models when running the protocol in Eqs.(\ref{eq:h3})-(\ref{eq:h1}), we found no continuous variation of the angles across the phase diagram of the kind, e.g., in Ref.~\onlinecite{zhou_quantum_2020}. However, we found that the optimal angles $\theta_{j,1}$ had a striking tendency to be very close to multiples of $\frac{\pi}{2}$ (see Fig.~\ref{fig:angle_distribution} in the Appendix). This suggests that the Hamiltonian $H_1$ has a restricted role in the evolution, and that the symmetry which led us to use a 3-step protocol could perhaps be broken in a simpler way. This property could be exploited by having the initial guess be close to multiples of $\frac{\pi}{2}$ through an ansatz, or by giving higher weight to regions close to these two points ($0$ and $\frac{\pi}{2}$) in the minimization algorithm.

\begin{figure}[ht]
	\centering
	\includegraphics{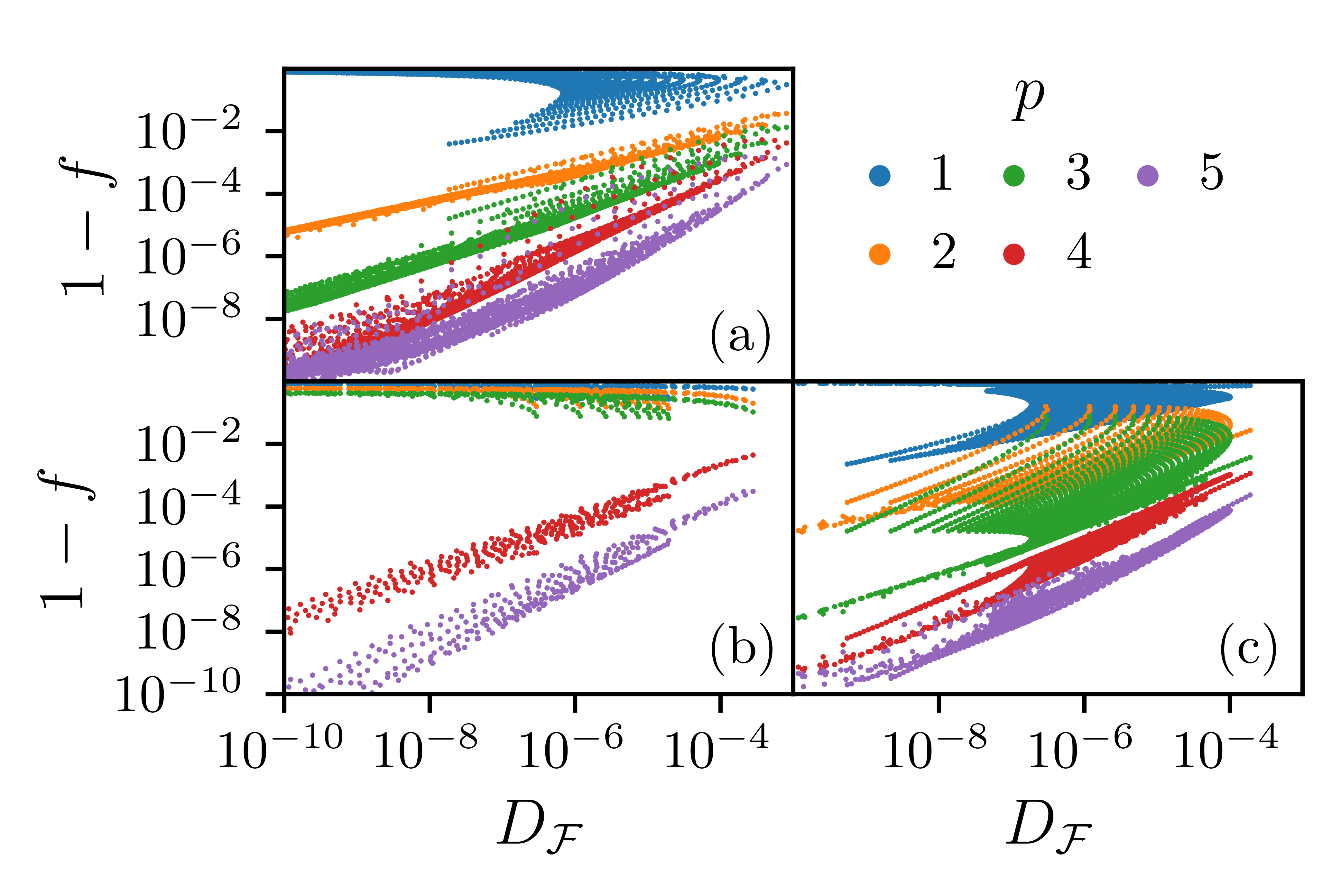}
	\caption{(a) Scatter plot of interaction distance $D_\mathcal{F}$ against QAOA's $1-f$ for the FM Ising model from Fig.~\ref{fig:4hmap}(c). (b)-(c): Analogous plot for the AFM Ising model in Fig.~\ref{fig:4hmap}(d), where the data points are taken either below (b) or above (c) the critical line. Data is for $N=8$ spins with PBCs and different values of $p$ indicated in the legend.}
	\label{fig:p_ising}
\end{figure}

In Fig.~\ref{fig:p_ising}, we show a scatter plot of $\df$ vs. $1-f$ from the data extracted from phase diagrams such as in Fig.~\ref{fig:4hmap}, but using different numbers of QAOA steps $p$, indicated in the legend. In both FM and AFM models, we expect correlation between $D_\mathcal{F}$ and $1-f$ around $p = \frac{N}{2}$. This correlation peak is relatively broad as $p$ is increased further. Eventually, as $p \to \infty$, we expect states to be exactly prepared and this correlation to break down, as in this limit our QAOA protocol should have the same power as quantum annealing with an arbitrary schedule~\cite{mbeng_optimal_2019}. In the opposite limit, as $p \to 1$, we expect that the variational method, in general, is not powerful enough for a correlation to emerge. However, in special cases such as the FM model, we see that $\df$ and $1-f$ are correlated even at lower $p$.  
We compute the Pearson correlation coefficients for the data in Fig.~\ref{fig:p_ising} and plot them in Fig.~\ref{fig:corr} of the appendix; as expected, the Pearson coefficient jumps to a value close to 1 around $p = \frac{N}{2}$.

For the AFM Ising model in Fig.~\ref{fig:p_ising}(b)-(c), we found that the correlation between $D_\mathcal{F}$ and $1-f$ follows a different slope in the two phases of the AFM model separated by the critical line in Fig.~\ref{fig:4hmap}(d). In particular, the behavior in the ordered phase of the model, Fig.~\ref{fig:p_ising}(b), clearly illustrates that the correlation between $D_\mathcal{F}$ and $1-f$ only starts to emerge around $p=N/2$. Moreover, different slopes of the correlation in the two phases suggest that by carefully examining the correlation between these two metrics one could infer about the existence of different phases in models with unknown phase diagrams. 

\section{Relation between QAOA and interaction distance}\label{sec:relation}

In Sec.~\ref{sec:ising_model}, we have established numerically a  correlation between $D_\mathcal{F}$ and the success of QAOA protocols. This suggests that the protocol's success depends on how close to being Gaussian (in the sense of Eq.~\eqref{eq:gaussian_manifold}) the target ground state is. In this section, we support these numerical observations by analytic arguments.

As mentioned in Sec.~\ref{sec:ising_model}, the angle $\theta_{j,1}$ associated with the operator Eq.~\eqref{eq:h3} is found to cluster around either $0$ or $\frac{\pi}{2}$. Now, note that a shift of $\pi/2$ in the $\theta_{j,1}$ part of the protocol results in an overall parity flip, as easily seen from the following sequence of identities:
\begin{align*}
	& \exp \left( i\left( \theta_{j,1} + \frac{\pi}{2} \right) \sum_i Z_i \right) \exp \left( i \theta_{j,2} \sum_i X_i \right) \\
	&= \exp \left( i  \theta_{j,1}  \sum_i Z_i \right) \prod_i Z_i  \exp \left( i \theta_{j,2} \sum_i X_i \right) \\
	&= \exp \left( i \theta_{j,1}  \sum_i Z_i \right) \exp \left( i \theta_{j,2} \sum_i - X_i \right) \prod_i Z_i.
\end{align*}
Further,
\begin{align*}
	\prod_i Z_i \ket{\rightarrow ... \rightarrow} = \ket{\leftarrow ... \leftarrow},
\end{align*}
where $|\rightarrow\rangle$, $|\leftarrow\rangle$ denote eigenstates of $X$.
This implies that, if we have the freedom of choosing either $\ket{\rightarrow ... \rightarrow}$ or $\ket{\leftarrow ... \leftarrow}$ as the initial state, we can restrict, without loss of generality,  all angles $\theta_{j,1}$ to an interval of length $\pi/2$. Since these angles are clustered around $0$ and $\pi/2$ as shown in Fig.~\ref{fig:angle_distribution} of the Appendix, they can all be mapped to be close to 0. 

Next, note that since both initial states are product states, they are also Gaussian states. Moreover,  the evolution under the unitaries generated by $H_3$ and $H_2$ maps Gaussian states into Gaussian states, while the evolution under $H_1$ spoils this property.  However, for $\theta_{j,1}$ close to $0$, the evolution under $H_1$  introduces only a small, perturbative deviation from a Gaussian state. This heuristically accounts for the high correlation of QAOA success with interaction distance of the target state, as the  states prepared by QAOA are close to being free. As $p$ gets larger, more perturbations are possible and the success of the QAOA increases. At a fixed $p$, the success of QAOA is related to the distance of the target state from the Gaussian state manifold. 

We conclude that there is a practical limitation to the ``natural" QAOA protocol proposed in Sec.~\ref{sec:ising_model}, which was obtained as a Trotter splitting of the Hamiltonian in Eq.~(\ref{eq:ising_hamiltonian}) into its translation invariant components: the protocol is unable to prepare ground states that are far from being Gaussian (as measured by interaction distance).
This limitation is fundamentally related to the probability spectrum of the target state, i.e., the eigenvalue spectrum of its reduced density matrix. Indeed, when performing QAOA using as a cost function the relative entropy~\cite{nielsen_quantum_2010} between the probability spectra of the trial state and of the target state, one finds heat maps similar to those in Fig.~\ref{fig:4hmap} (data not shown). Thus, there is a correlation between QAOA and $\df$, even though  the former minimizes the overlap of two vectors, while the latter employs a minimization using the probability spectrum of the subsystem's reduced density matrices.

\begin{figure}[t]
	\centering
	\includegraphics{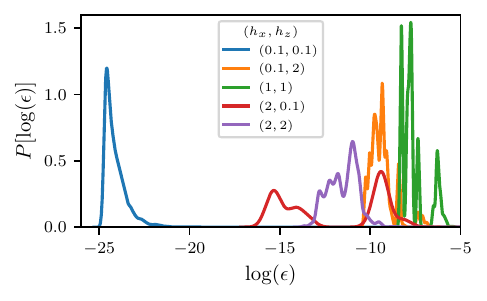}
	\caption{Probability distribution function $P(\log(\epsilon))$ of QAOA outcomes on 10000 uniformly generated initial angles having as target the ground state of the points in $S$. Data is for system size $N=8$ and $p=4$, with the protocol in Eqs.~\eqref{eq:h1}-\eqref{eq:h3}.
	}
	\label{fig:epsilon_distrib}
\end{figure}

\begin{figure*}
	\centering
	\includegraphics{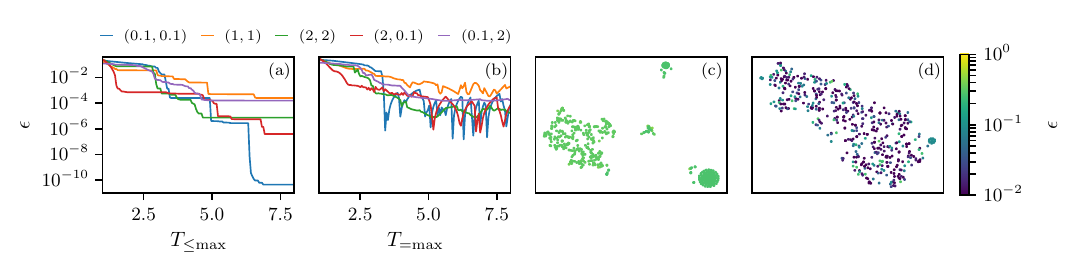}
	\caption{Minimization landscape in the AFM Ising model.
	(a) Relative energy $\epsilon$ vs. $T_{\leq \max}$ for different points in $S$.  (b) Relative energy $\epsilon$ vs. $T_{= \max}$ for different points in $S$. (c) t-SNE graph for 500 random angle samples taken at $(h_x, h_z) = (1,1)$ and $T_{= \max} = 1$. (d) t-SNE graph for 500 random angle samples taken at $(h_x, h_z) = (1,1)$ and $T_{= \max} = 8$. All data is for system size $N=6$ and $p=3$. Color scale in (c), (d) represents the value of $\epsilon$.}
	\label{fig:4line_landscape}
\end{figure*} 

\section{Minimization landscape}
\label{sec:landscape}

In this section, we explore the  minimization landscape of the optimization problem studied in Sec.~\ref{sec:ising_model} for the AFM Ising model (we reached qualitatively similar conclusions in the FM model). The target state in the cost function is taken to be the ground state of the Hamiltonian at a set of representative points in the ($h_x$, $h_z$) phase diagram, $S = \{(0.1, 0.1), (0.1, 2), (1, 1), (2, 0.1), (2, 2)\}$. These points are drawn from  regions of both ``hard" and ``easy" state preparation according to Fig.~\ref{fig:4hmap}. Here, we use the rescaled relative energy, defined in Eq.~(\ref{eq:relative_energy}), instead of the quantum infidelity.

In Fig.~\ref{fig:epsilon_distrib} we first look at the probability distribution function for $\log\epsilon$ in selected points $S$.
We generate a sample of $10^4$ initial $\theta$ angles, drawn from a uniform distribution in the $[0,\pi[$ interval.  The distribution of $\log\epsilon$ gives us insight about the structure of the landscape. A sharply-defined distribution of $\log \epsilon$ is only obtained in the case where $h_x=h_z=0.1$, with the peak at $\epsilon$ close to 0. The mean of the distribution shifts to large values of $\epsilon$ upon approaching the critical line, e.g., at $h_x=h_z=1$. In addition to the shift of the mean, the distribution also develops multiple peaks corresponding to local minima.  At other points in the phase diagram, such as $h_x=h_z=2$, the separate minima form a smooth curve with larger variance. Finally, in some cases like $h_x=2$, $h_z=0.1$, we observe a clear bimodal distribution of the minima.  Thus,  the distribution of minima varies considerably across the phase diagram and, generally, has multiple peaks.

A systematic investigation of the nature of the landscape of a related minimization problem was performed in Refs.~\onlinecite{bukov_broken_2018, day_glassy_2019,liang_investigating_2020} using a discretized adiabatic state preparation protocol. In these works, the behavior of the minimization landscape was examined as a function of the total allowed time for the protocol. It was found~\cite{bukov_broken_2018, day_glassy_2019} that there are distinct ``phases" associated with different intervals for the total allowed time. Particularly, at intermediate times, there is a glassy phase presenting with multiple clusters of minima where the minimization becomes difficult. Following Refs.~\onlinecite{bukov_broken_2018, day_glassy_2019},  we have probed the nature of the minimization landscape in our models and using our QAOA protocol when the total time $T({\bm{\theta}}) = \sum_{i=1}^p \sum_{ j=1}^{M} \theta_{i, j}$ is restricted. We impose this restriction in two different ways. First, we allow $T({\bm{\theta}})$ to be less than or equal to some maximum total time $T_{\leq\max}$, which can be easily achieved by constraining the allowed interval for each $\theta_{i,j}$ angle in our protocol. The second method is to demand $T({\bm{\theta}})$ to be exactly equal to a given total time $T_{=\max}$. The results of these two approaches are contrasted in Fig.~\ref{fig:4line_landscape}(a) and (b). 

In Fig.~\ref{fig:4line_landscape}(a) we see that, as expected, as $T_{\leq \max}$ increases, $\epsilon$ decreases. Perhaps surprisingly, this occurs in a very clear step-wise fashion, suggesting that there are discrete values of $T_{\leq \max}$ that show significant improvement in state preparation.  By contrast, in Fig.~\ref{fig:4line_landscape}(b) we see that as $T_{= \max}$ increases, the behavior of $\epsilon$ is more erratic, indicating that there are discrete, optimum values of $T$ for which states can be prepared under this restriction. This shows that the protocol can not accommodate non-optimal values $T_{= \max}$, that is, there is no way for the protocol to continuously "stall" and wait, "wasting time" so as to emulate the last optimal value of $T_{= \max}$. The protocol can, however, "stall" in discrete values of $\pi$, due to the symmetry in the angles. A consequence of this seems to be the peaks and troughs pattern in the graphs in Fig.~\ref{fig:4line_landscape}(b), which show an irregular pattern.  This contrasts with the results in Ref.~\onlinecite{bukov_broken_2018, day_glassy_2019}, which display an almost monotonically increasing success in state preparation as $T_{= \max}$ increases. 

Next, we took 500 random angle samples restricted to either $T_{= \max}$ or $T_{\leq \max}$ and ran QAOA with target state coming from the ground state at each of the representative points in $S$. Here, we have used the L-BFGS local optimization algorithm, as implemented in the \texttt{scipy} Python package, to perform the minimization.
In order to plot the high-dimensional minimization landscape, we have used t-SNE~\cite{maaten_visualizing_2008}, a dimensionality-reduction algorithm for data visualization that embeds high dimensional data in a space with lower dimension while preserving the relative position of the data points. Performing t-SNE on these samples, we find that, for $T_{\leq \max}, T_{= \max} < 1$, there exists clustering of minima, although some of the clusters are significantly less compact than others -- see  Fig.~\ref{fig:4line_landscape}(c).
For $T_{\leq \max}, T_{= \max}>1$, the clustering rapidly disappears, first for the $T_{\leq \max}$ restriction and then for the $T_{= \max}$ restriction -- an example of the latter is shown in Fig.~\ref{fig:4line_landscape}(d) for 
$T_{= \max}=8$. This indicates that QAOA, which usually does not place restrictions on the values of $\theta$ angles and therefore implicitly operates in the large-$T$ regime, does not display a glassy phase in its minimization landscape as found for a different protocol in Refs.~ \onlinecite{bukov_broken_2018, day_glassy_2019}.

\section{Conclusions}
\label{sec:conclusion}

In this paper we have investigated the preparation of ground states of non-integrable quantum models using QAOA. Our motivation was to identify physical properties of the state that have an impact on  its preparation, thereby allowing us to bound the relative success of QAOA. While this task appears challenging for rigorous analytical treatment, we have numerically demonstrated a correlation between interaction distance and the success of the QAOA protocols in several variants of the quantum Ising model. This suggests that, in these models, states which are far from free, as measured by interaction distance, are harder to prepare, i.e., in order to prepare states with larger interaction distance, QAOA needs higher values of $p$ to achieve the same degree of success as for states with lower interaction distance and lower $p$. We have also performed an analysis of the landscape associated with the QAOA optimization problem. We have found that there are several local minima associated with this landscape, though they are spread out and show no distinctive clustering. Limiting the total allowed QAOA time did not alter this landscape significantly for total time $T\gtrsim 1$. 

One of the applications of our results is that theoretical insight into the closest free states representing the target state can be gained by using the experimentally obtained QAOA ansatz and setting the small $\theta_{j,1}$ angles to be zero. The absence of the glassy phase in the minimization landscape implies that the natural QAOA protocols constructed here do not lead to a NP-hard optimization problem and the time to find optimal angles should scale polynomially with the system size.

\section{Acknowledgements}

We acknowledge useful discussions with Chris Self, Christopher Turner, and Marin Bukov. This work was supported by EPSRC grant EP/R020612/1. Statement of compliance with EPSRC policy framework on research data: This publication is theoretical work that does not require supporting research data. 

\bibliography{df_qaoa}

\appendix

\section{Three-spin Ising model}
\label{sec:three-spin}

Here we demonstrate that our findings from the main text also apply in a different model featuring three-spin interactions. The model is defined by the Hamiltonian
\begin{align} \label{eq:3ising_hamiltonian}
	H &= - \sum_{i = 1}^{N} Z_i Z_{i+1} Z_{i+2} - h_x \sum_{i = 1}^{N} X_i - h_z \sum_{i = 1}^{N} Z_i ,
\end{align}
where, again, $X_i$ and $Z_i$ are the standard Pauli matrices on site $i$, and we assume PBCs. The critical behavior of this  model is in the same universality class as the two-dimensional classical three-state Potts model. The phase diagram of the model has been mapped out in Ref.~\onlinecite{penson_conformal_1988} (see also Ref.~\onlinecite{igloi_quantum_1989} for further generalisations of the model). In the $h_x > 0$, $-3 \leq h_z < 0$ region, it contains a critical line connecting $(h_x, h_z) = (0, -3)$ to $(h_x, h_z) = (1, 0)$. Below that critical line, there exists a threefold ground state degeneracy, while above it the ground state is unique. 

\begin{figure}[ht]
	\centering
	\includegraphics[width=\linewidth]{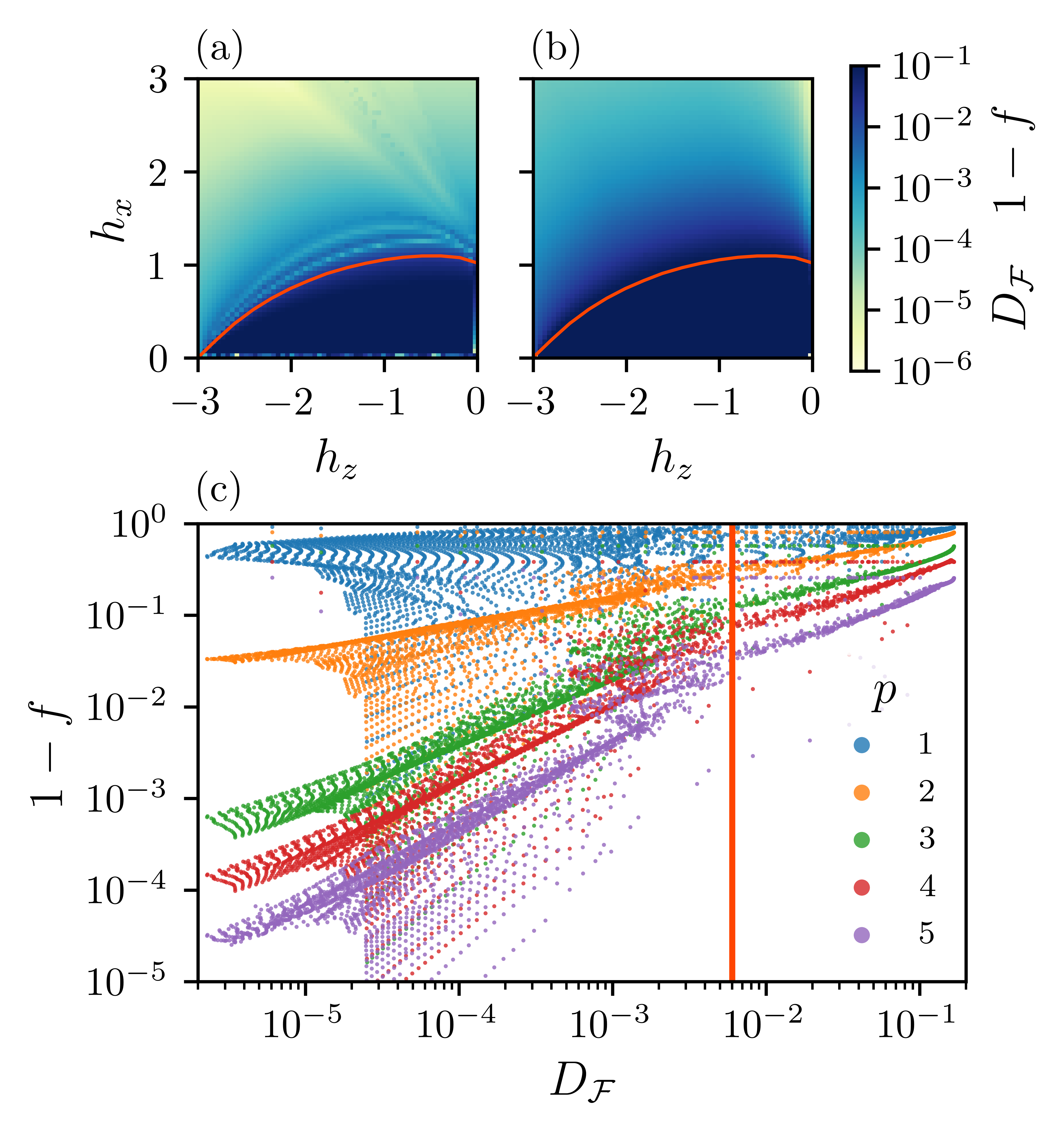}
	\caption{Top row: Interaction distance for the 3-spin Ising model in Eq.~(\ref{eq:3ising_hamiltonian}) as a function of fields $h_x$ and $h_z$ (a) and the infidelity $1-f$ of the QAOA protocol (b). The QAOA protocol is based on Eqs.~\eqref{eq:h3}, \eqref{eq:h2}, \eqref{eq:h1_3ising}. 
	All data is for system size $N=9$ using PBCs. Red line is the critical line in the 3-spin Ising model according to Ref.~\onlinecite{penson_conformal_1988}
Bottom row: Scatter plot of $D_\mathcal{F}$  against $1-f$ for the 3-spin Ising model, system size $N=9$ and different values of $p$ indicated in the legend. Vertical red line, drawn heuristically, separates the data belonging to the phase above and below the critical line (respectively to the left and right of the vertical line). 
}.
\label{fig:3ising_heatmaps}
\end{figure}

The motivation for studying the model in Eq.~(\ref{eq:3ising_hamiltonian})
is that its ground state is expected to be more strongly interacting and have a higher value of $D_\mathcal{F}$. For example, unlike the FM/AFM models in Eq.~(\ref{eq:ising_hamiltonian}), the ground state of the model in  Eq.~(\ref{eq:3ising_hamiltonian}) cannot be obtained in closed form using the Jordan-Wigner and/or Bogoliubov transformation (apart from classical line, $h_x=0$). Moreover,  the 3-fold ground state degeneracy in the ordered phase gives rise to an approximate 3-fold degeneracy of the entanglement spectrum, as generally found in ``symmetry-protected topological phases"~\cite{pollmann_entanglement_2010}. This can be understood by picking a point $(h_x=0, h_z=-1)$, where the exact ground state of the system (with zero momentum under translation) is given by
\begin{align}\label{eq:3gs}
 |\psi_0\rangle = \frac{1}{\sqrt{3}}( |\uparrow \downarrow \downarrow \uparrow \downarrow \downarrow \ldots\rangle +  |\downarrow \uparrow \downarrow \downarrow \uparrow \downarrow \ldots\rangle 
+ |\downarrow \downarrow \uparrow \downarrow \downarrow \uparrow \ldots\rangle ). \quad\quad
\end{align}
The corresponding entanglement spectrum is given by $\rho_k = \{ \frac{1}{3}, \frac{1}{3}, \frac{1}{3}, 0,0, \ldots \}$. This is the type of entanglement spectrum that gives $D_\mathcal{F}=\frac{1}{6}$, a value close to the upper bound $3-2\sqrt{2}$~\cite{meichanetzidis_free-fermion_2018}. The same spectrum is obtained in the $\mathbb{Z}_3$ parafermion model at its fixed point~\cite{mong_parafermionic_2014}. An approximate 3-fold degeneracy in the entanglement spectrum persists throughout the ordered phase of the model, thus we expect the ground state throughout this phase to be more difficult to  prepare using QAOA compared to the disordered phase. 

The comparison between $D_\mathcal{F}$ and QAOA for the model in Eq.~(\ref{eq:3ising_hamiltonian}) is shown in Fig.~\ref{fig:3ising_heatmaps}. The QAOA protocol was chosen such that $H_1$ and $H_2$ are defined as in Eqs.~(\ref{eq:h3})-(\ref{eq:h2}), but for $H_3$ we use
\begin{align} \label{eq:h1_3ising}
	H_3 = - \sum_{i = 1}^{N} Z_i Z_{i+1} Z_{i+2}.
\end{align}
Note that this protocol also satisfies $H_3+h_xH_2+h_z H_1=H$. We have found that, like the two-spin Ising model, the success of the protocol also correlates well with interaction distance, as we see in the top row of Fig.~\ref{fig:3ising_heatmaps}. Here, as in Section~\ref{sec:ising_model}, minimizations are done using a basinhopping algorithm, and the results required two additional rounds of minimization. Moreover, we find correlation between $D_\mathcal{F}$ and $1-f$ for several values of $p$, as shown in the bottom row of Fig.~\ref{fig:3ising_heatmaps}. As before, the data in the bottom row of Fig.~\ref{fig:3ising_heatmaps} was obtained by sampling across the entire phase diagram in the top row of Fig.~\ref{fig:3ising_heatmaps}.

It worth noting that we can prepare the ground state in Eq.~(\ref{eq:3gs}) exactly by choosing the protocol $H_2 = - \sum_j Z_{j-1}Z_jZ_{j+1}$ and $H_1 =-\sum_j(X_jX_{j+1}+Y_jY_{j+1})$, while the initial state is the ground state of $H_2$ in the sector with magnetization $-N/3$, as this is the sector where the states $\{\ket{\uparrow \downarrow \downarrow...}, \ket{\downarrow \uparrow \downarrow...}, \ket{\downarrow \downarrow \uparrow...}\}$ live. It can be verified that this protocol prepares the exact ground state in Eq.~(\ref{eq:3gs}) in $N/2$ steps. Moreover, supplementing the protocol with a third operator, $H_3 = -\sum_j X_j$, leads to good results across the entire phase with the 3-fold ground-state degeneracy. However, the infidelity $1-f$ of the latter protocol does not capture the phase transition in a way that the protocol [Eqs.\eqref{eq:h2}, \eqref{eq:h3}, \eqref{eq:h1_3ising}] does. Moreover, the initial state is more difficult to prepare in this case, unlike the product state of spins in our protocol.

Similar to the models studied in Sec.~\ref{sec:ising_model}, we found no continuous variation of angles in the three-spin Ising model, and the angles  $\theta_{j,1}$ tended to be close to multiples of $\pi/2$ (see Fig.~\ref{fig:angle_distribution}). However, in this case the heuristic arguments of  of Sec.~\ref{sec:relation} do not directly apply as the Gaussianity of the protocol is broken by the triple spin interaction term~(\ref{eq:h1_3ising}). It is an interesting open problem to analytically explain the approximate Gaussianity in this case.

\section{Cluster Ising model}\label{sec:cluster}

\begin{figure}
    \centering
    \includegraphics[width=0.97\columnwidth]{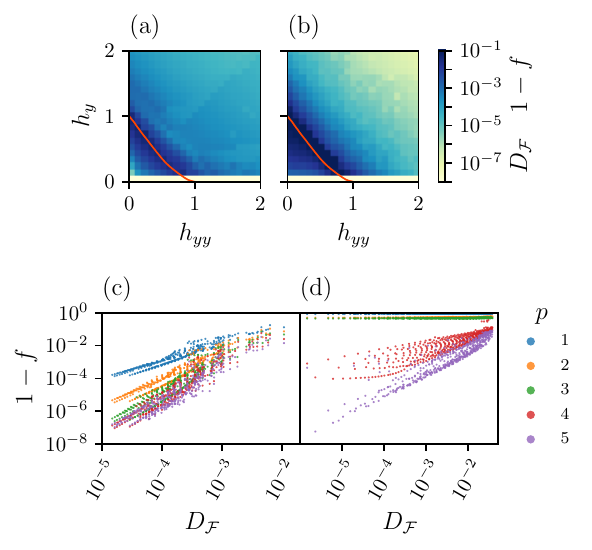}
    \caption{
    Top row: Interaction distance for the cluster Ising model in Eq.~(\ref{eq:cluster}) as a function of $h_{yy}$ and $h_y$ (a) and the infidelity $1-f$ of the QAOA protocol (b).
All data is for system size $N=8$ with periodic boundary conditions. Red line is reproduced from Ref.~\onlinecite{verresen_one-dimensional_2017}. Bottom row: Scatter plot of $D_\mathcal{F}$ against $1-f$ for the cluster Ising model, system size $N=8$ and different values of $p$ indicated in the legend. Panel (c) shows points above the critical line while (d) shows points below it.
}
    \label{fig:cluster_ising}
\end{figure}
As another example illustrating the versatility of our method, we consider the cluster Ising model, which has recently attracted attention both in the context of quantum information~\cite{pachos_three-spin_2004, smacchia_statistical_2011} and symmetry-protected topological phases~\cite{verresen_one-dimensional_2017}. The model is defined in terms of Pauli matrices (assuming periodic boundary conditions) as: 
\begin{eqnarray}\label{eq:cluster}
    H = - \sum_{i=1}^N X_{i-1} Z_{i} X_{i+1} - \sum_{i=1}^N h_{yy} Y_i  Y_{i+1} +  \sum_{i=1}^N h_y Y_i. \quad \quad
\end{eqnarray}
When the field $h_y=0$, the model can be solved using a combination of Jordan-Wigner/Bogoliubov transformations~\cite{smacchia_statistical_2011}, but for general values of $h_y$ the model is not solvable. The model has a critical line described by conformal field theory with central charge $c=3/2$, connecting points $(h_{yy}=1,h_y=0)$ and $(h_{yy}=0,h_y=1)$. The critical line has been mapped out using density-matrix renormalisation group calculations in Ref.~\onlinecite{verresen_one-dimensional_2017}.

As seen in Fig.~\ref{fig:cluster_ising}(a)-(b), both  QAOA and $D_\mathcal{F}$ are highly sensitive to the critical line, just like we have previously seen in the AFM Ising and 3-spin Ising models. Despite small system size, the critical behavior is in good qualitative agreement with results of Ref.~\onlinecite{verresen_one-dimensional_2017}. The QAOA protocol in Fig.~\ref{fig:cluster_ising} has been defined by splitting $H$ into its three components, $H_3 = - \sum_{i=1}^N X_{i-1} Z_{i} X_{i+1}$, $H_2 = -  \sum_{i=1}^N Y_i Y_{i+1}$, $H_1 = -  \sum_{i=1}^N Y_i$. For the initial state, one can choose the ground state of $H_1$. However, with this initial state, the convergence of the optimization below the critical line was found to be very slow; instead, the convergence is considerably more robust if we use as initial state the ground state of $H_2$ in this regime. In producing the phase diagram in Fig.~\ref{fig:cluster_ising}(b) we have run two sweeps of QAOA starting in either of these initial states, and plotting the smaller value of the obtained $1-f$.  

It is worth noting that the line with $h_y=0$ is prepared \emph{exactly} (to machine precision) in $p=N/2$ steps using the 2-step protocol involving only $H_3$ and $H_2$, similar to the case of the transverse field Ising model. Moreover, there is correlation between $D_\mathcal{F}$ and $1-f$ across each of the two phases of the model, as illustrated in Fig.~\ref{fig:cluster_ising}(c)-(d).

\section{Additional data}

\begin{figure}
		\centering
		\includegraphics[width=\linewidth]{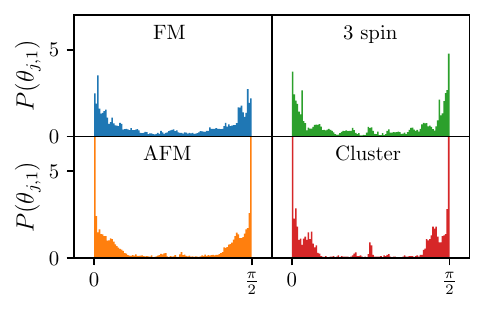}
		\caption{Distribution of angles $\theta_{j,1}$ associated with the Hamiltonian in Eq.~\eqref{eq:h3} in QAOA evolution across phase diagrams of FM/AFM, 3-spin and cluster Ising models in Eqs.~(\ref{eq:ising_hamiltonian}), Eq.~(\ref{eq:3ising_hamiltonian}) and Eq.~(\ref{eq:cluster}). Data is for system size $N=8$ with the exception of $N=9$ for the 3-spin Ising model. In all the plots, $p=4$.}
		\label{fig:angle_distribution}
\end{figure}
This section contains some additional results that support the conclusions in the main text.  Fig.~\ref{fig:angle_distribution} shows the distribution of the angles $\theta_{j,1}$ associated to the ``interacting" part of the Hamiltonian, i.e., $H_1$ in Eq.~(\ref{eq:h3}) in the QAOA protocol. As claimed in Sections~\ref{sec:ising_model}-\ref{sec:relation}, these angles tend to be close to $0$ or $\frac{\pi}{2}$, which is clearly seen in the figure for all the models considered in this paper.

\begin{figure}[ht]
    \centering
    \includegraphics{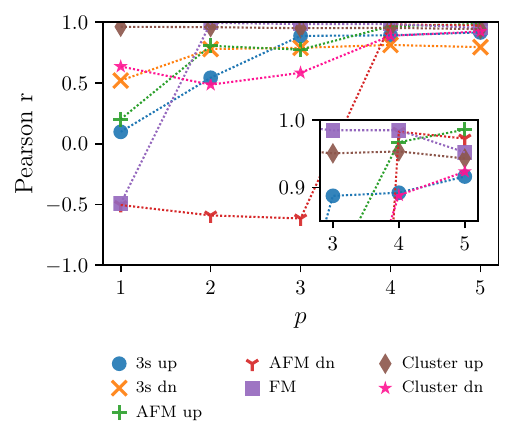}
    \caption{Pearson correlation coefficients as a function of $p$ for the data in Figures~\ref{fig:p_ising},~\ref{fig:3ising_heatmaps}(c) and ~\ref{fig:cluster_ising}(c). Labels "up" and "dn" denote data above and below the critical line of the corresponding model, respectively. The inset zooms in on the top-right corner of the plot.}
    \label{fig:corr}
\end{figure}
Finally, we compute the Pearson correlation coefficients for the data in Figs~\ref{fig:p_ising},~\ref{fig:3ising_heatmaps}(c) and ~\ref{fig:cluster_ising}(c)-(d), and plot them in Fig.~\ref{fig:corr}. 
We see that the Pearson coefficient jumps to a value close to 1, indicating direct correlation. As explained in the main text, we expect this to mark the beginning  of a broad plateau where the Pearson coefficient remains close to 1, until it eventually starts to drop at larger values of $p$. The reason for this decay is the exact preparation of the state in the limit $p\to\infty$ for the protocol considered here. Conversely, in the limit $p\to 1$, we expect the variational ansatz is not sufficiently powerful for the correlation to emerge. 

\end{document}